\documentclass[preprint]{revtex4-1}
\usepackage{epsf,epsfig}
\usepackage[psamsfonts]{amssymb}
\usepackage{bm}
\usepackage{graphicx}
\usepackage{dsfont}
\usepackage{amsmath}
\usepackage{natbib}
\usepackage{xcolor}
\usepackage{mathrsfs}
\usepackage{xcolor}
\newcommand{\be}{\begin{equation}}
\newcommand{\ee}{\end{equation}}

\newcommand{\bea}{\begin{eqnarray}}
\newcommand{\eea}{\end{eqnarray}}
\newcommand{\nn}{\nonumber}
\newcommand{\commentold}[1]{}
\begin{document}
\title{Propagation and Dynamics of Oscillating Tetraquark Systems}
\author{Sh. Janjan}
\email{sh.janjan@razi.ac.ir}
\affiliation{Department of Physics, Razi University 67149, Kermanshah, Iran}
\author{G.R. Boroun}
\email{boroun@razi.ac.ir}
\affiliation{Department of Physics, Razi University 67149, Kermanshah, Iran}
\date{\today}
\begin{abstract}
\noindent{ We have presented a theoretical investigation into the behavior of tetraquark states. These states are modeled as a non-relativistic four-body quantum system governed by harmonic interactions. Our goal is to  capture the fundamental dynamics of quark-antiquark interactions in the tetraquark configuration.  Within this framework, we  have derived an analytical expression for the quantum propagator. This expression was then used to compute the time evolution of the wave function and the associated probability density. We examined the time evolution and oscillation patterns to gain a better understanding of the wave function and its probability of presence in space.  This approach offers valuable insights into the dynamics of the tetraquark system. This model can serve as a basis for predicting complex structures such as entanglement and multipartite dependence of quarks.}
\end{abstract}
%
%
\keywords{}
\date{\today}
\maketitle
%
\section{Introduction}\label{Introduction}
\noindent In particle physics, tetraquarks are particles made up of four quarks, tetraquarks are complex structures  consisting of four quarks and antiquarks. These particles are less familiar than others in the filed, so a better understanding of their properties would significantly enhance  our knowledge and scientific progress in particle physics \cite{R0001,R0002,R0003}. In 2003, the tetraquark $X(3872)$ was discovered by the Bell collaboration, which reported a candidate for a narrow charmonium-like tetraquark. Also of note is $X(6900)$, which was interpreted as a full-charm tetraquark \cite{R001,R002}. In 2013, the tetraquark $Z_{c}(3900)$ was experimentally discovered for the first time by two independent experiments, confirming theoretical predictions about the existence of such particles. Additionally $T^{+}_{cc}$, was observed in 2021 by the LHCb, marking the first observation of a stable excited tetraquark with two heavy quarks. These discoveries are major milestones in the exploration of multiquark states and contribute to the understanding of multiquark dynamics. Recently, many researchers have delved into tetraquark systems and their characteristics, making this area of study a continuing interest in current reseach \cite{R02,R03,R04,R05,R06,R07,R008}.

 The motion of particles with harmonic oscillations is one of the most important problems in physics, such as the oscillations of atoms in nuclei and the harmonic oscillator problem. Moreover, the quantum theory of electromagnetic fields is somehow related to the problem of harmonic oscillators \cite{R1,R2,R3,R4}. When a particle is in a strong potential field,  relativistic effects must be taken into account. By solving theoretical problems such as diffraction, spectroscopy, interference, and dynamics of particle motion, we can make significant contributions to these issues. To study the dynamics of particles and investigate their motion paths, we use  tools such as a propagator to determine the path of the particle and the probability of its presence at later times \cite{R5,R6,R62}. However, in relativistic and non-relativistic quantum mechanics, many authors pay special attention to the harmonic oscillator problem.

 Quarks are fundamental particles that serve as the building blocks of matter. They are never found individually due to color confinement, and  can be located inside hadrons where they are bound together by gluons. Therefore, in much scientific research concerning fundamental particles, multi-quark systems such as tetraquarks, pentaquarks, etc. hold significant importance \cite{R7,R71,R8,R9,R10,R101}. The study of quark oscillations is also crucial, as understanding the motion of particles like quarks with harmonic oscillations is a fundamental problem in physics,  that  has been extensively explored in various articles.

 In 1971, Richard Feynman developed a relativistic equation to describe the symmetric quark model with harmonic interaction. He calculated the decay width and flow matrix elements, all of which were of first order \cite{R11}. Additionally, Sonia Kabana and Peter Minkowski studied the oscillation modes of three-quarks \cite{R12}. Various methods exist for solving many-body systems and the corresponding Schrödinger equation \cite{R13}. For instance, the Gaussian expansion method has been used to solve the Schrödinger equation and calculate the mass spectrum of heavy tetraquark systems. It has been observed that the kinetic energy of tetraquarks resembles that of molecular states \cite{R14}.  The propagator appears in the Feynman diagram as lines that connect virtual particles between interaction points, connecting particles at the vertices \cite{R5}.

  In 2012, M. K. Bahar and F. Yasuk used the (AIM) method to solve quadratic equations by considering the harmonic oscillator potential.   They determined the eigenvalues of relativistic energy and  the corresponding functions for the harmonic oscillator interaction potential and the quark-antiquark interaction \cite{R4}. In these works, the Gaussian wave function were used as a basis, and the spatial wave function of the tetraquark was considered  a Gaussian function as well \cite{R16}. Additionally, in 2022, the ground state of heavy tetraquarks was investigated by solving nonrelativistic four-body systems. Both butterfly and flip-flop configurations were studied in Ref.\cite{R17}. The study of quark motion and their spatial changes overtime is crusial. By observing the motion of quarks, one can analyze the properties of the wave that reaches them. For example,  gravitational wave signals cause quarks to oscillate, generated by merging black holes or colliding neutron stars \cite{R18}.

   The Hamiltonian used in our work consists of three terms: potential energy, kinetic energy, and momentum \cite{R19}. Additionally, in the work of 2023, a new Hamiltonian was applied to attractive tetraquarks,  resulting in the discovery of a bound state and two resonances \cite{R20}. In many studies on the interaction between heavy quarks,  researchers use the short-range gluon exchange potential instead of the long-range potential \cite{R16}. In various scenarios, the harmonic oscillator potential is considered for tetraquark problems due to the system$^,$s oscillations in real multi-quark systems, allowing for the investigation of its analytical solution \cite{R21}.

   In 2024, Benoît Assi and his colleagues obtained the Green's function of the ground state energies of the tetraquark using the Monte Carlo method. The wave function used in this work was treated more as an experimental state, and a set of experimental wave functions
   were investigated, which are crucial in studying the dynamics of tetraquarks \cite{R212}. It is important to note that the dynamics of quarks in moving oscillatory states were discussed, and the oscillation frequency of quark motion in the ground state was also studied. When  particles are placed in a strong field, relativistic effects become significant, as studied in literature \cite{R22,R23,R24}. Due to the complexity of the dynamics of these systems, most problems have not addressed it,  and no specific order has been predicted for the dynamics of multi-quark models. This work  attempts to provide insight into the oscillations of the tetraquark system and its dynamics. Indeed,  It demonstrates  how a tetraquark system, as a many-body system, evolves over time.

 In the framework of QCD, inspired by QFT, we present an analytical model for the evolution of quarks in a tetraquark system. This model is a phenomenological and effective framework for understanding the structure and dynamics of tetraquarks. Using quantum methods, we will derive a wave function that captures the dynamical properties of the tetraquark system by considering all the parameters of the degrees of freedom. In phenomenological models for gluons confined in hadrons, such as tetraquarks, it is acceptable to model oscillatory behavior for gluons and to consider oscillatory behavior for gluon coupling. Hence, the model and structure presented in this work are based on the oscillatory behavior of gluons, where gluon fields exhibit wave-like or oscillatory behavior in the tetraquark Hamiltonian. The wave function obtained in this work provides a realistic representation of QCD, and this model can also help predict complex structures such as entanglement and multipartite dependence of quarks.

In Section \ref{H}, the Hamiltonian of the tetraquark system is introduced, and the propagator and wave function of  tetraquark systems are obtained. Section \ref{P}, focuses on obtaining the probability density and dynamics of tetraquark systems over time. Finally, in Section IV, the results of this research are discussed.

\section{Hamiltonian and Propagator of the Tetraquark System }\label{H}
\noindent The non-relativistic Hamiltonian of the multiquark system is written as follows:
\be\label{1}
H=\sum_{i=1}^{N}\Big(\frac{\mathbf{p}_{i}^2}{2\mathbf{m}_i}+\mathbf{m}_i\Big)-T_{cm}+\sum_{i<j=1}^{N}(V^{C}_{ij}+V^{G}_{ij}),
\ee
where $T_{cm}$ is the kinetic energy component of the center of mass and ${p}_i$ and $m_i$ represent the momentum and mass of  the ith quark or antiquark, respectively; that all quarks are considered non-relativistic. Introducing  parameters $V^{C}_{ij}$ and $V^{G}_{ij}$ associated with the spin-spin interaction \cite{R17,R25}, the Hamiltonian of the four interacting quark system can be written as:
\be\label{2}
H=\sum_{i=1}^{4}\Big(\frac{{\mathbf{p}}_i^2}{2 \mathbf{m}_i}+\mathbf{m}_i\Big)-T_{cm}+\sum_{i<j=1}^{4}\Big(-\overrightarrow{\lambda}_i.\overrightarrow{\lambda}_ja\mathbf{x}_{ij}^2+\alpha_s\frac{\overrightarrow{\lambda}_i. \overrightarrow{\lambda}_j}{4\mathbf{x}_{ij}}-\alpha_s\frac{\overrightarrow{\lambda}_i. \overrightarrow{\lambda}_j}{4}\frac{\pi\delta(\mathbf{x}_{ij})}{\mathbf{m}_i\mathbf{m}_j}(1+\frac{2}{3}\overrightarrow{\sigma_i}.\overrightarrow{\sigma_j})\Big),
\ee
where $\alpha_s$ is the quark-gluon coupling constant and $a$ is the strength of the confinement potential. Additionally, $\overrightarrow{\lambda}_i$ and $\overrightarrow{\sigma_i}$ represent the color generator and Pauli operator, respectively. Also $\mathbf{x}_{ij}=x_{i}-x_{j}$,which denotes the location of ith particle. For simplicity in calculations, we introduce new parameters as $\omega=\overrightarrow{\lambda}_i.\overrightarrow{\lambda}_j\,a$, $\beta=(\alpha_s/4)\overrightarrow{\lambda}_i. \overrightarrow{\lambda}_j$ and $\beta'=(\pi/\mathbf{m}_i\mathbf{m}_j)(1+2\overrightarrow{\sigma_i}.\overrightarrow{\sigma_j}/3)$.
 We consider the color-confinement interaction in quadratic form as $V^{C}_{ij}=-\overrightarrow{\lambda}_i.\overrightarrow{\lambda}_j\,a\,\mathbf{x}_{ij}^2$\cite{R26}. Therefore, the desired Hamiltonian will be a quadratic Hamiltonian. We can derive the propagator of a four-particle system with a quadratic Hamiltonian, where the particles  interact with each other in an oscillatory method. The Hamiltonian of our tetraquark system includes an oscillatory term $(\mathbf{p}_{i}^2/2\mathbf{m}_i)+\sum_{i<j=1}^{4}(-\overrightarrow{\lambda}_i.\overrightarrow{\lambda}_j\,a\,\mathbf{x}_{ij}^2)$.
For the tetraquark, we can consider Fig.(1), which is known as a flip-flop form. We set $N=4$ to consider a tetraquark system, where the problem involves $x_1$,$x_2$,$x_3$ and $x_4$,which are the positions of the tetraquarks.In fact, the value of $\sum_{i<j=1}^{4}\mathbf{x}_{ij}=\sum_{i<j=1}^{4}x_{i}-x_{j}=[(x_1-x_2)+(x_2-x_3)+(x_3-x_4)+(x_1-x_4)]$ and $\sum_{i=1}^{4}\mathbf{p}_i=p_1+p_2+p_3+p_4$.For systems with the tetraquark Hamiltonian, we can find the propagator using the Heisenberg approach for quantum systems, where momentum and position are time-dependent variables $p_{i}(t)$ and $x_{ij}(t)$, respectively \cite{R27}. In Hamiltonian Eq.(\ref{2}), the term $\sum_{i<j=1}^{4}(\beta/\mathbf{x}_{ij})-\beta\,\beta'\,\mathbf{x}_{ij}$ acts as an external force on the system, where we defind $\sum_{i=1}^{4}\mathbf{x}_{ij}=x_i$ and $\sum_{i=1}^{4}\mathbf{m}_i=m_i$; also $\sum_{i=1}^{4}\mathbf{p}_i=p_i$.
\begin{figure}\label{F1}
\centering
\includegraphics[scale=0.5]
{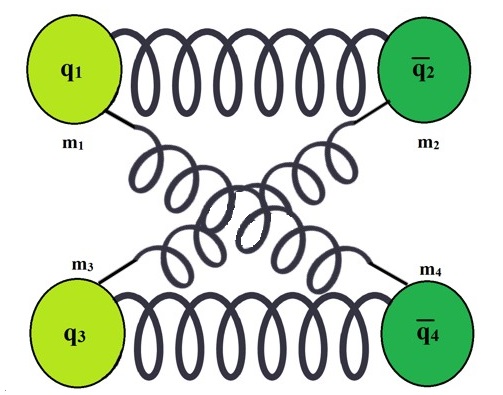}
\caption{Two quarks $q_1$ and $q_3$ and two antiquarks $\overline{q}_2$ and $\overline{q}_4$ are brought together in a tetraquark system with an oscillatory interaction. The position of each particle relative to the others is denoted by$\sum_{i<j=1}^{4}{x_{ij}}$ and the mass of each particle is denoted by $\sum_{i=1}^{4}m_i$.}
\end{figure}
Considering the tetraquark as an oscillating system moving from the initial state to the secondary state, the transition matrix that describes the probability distribution of the system at times of this system can be obtained as a $4\times4$ matrix with eigenvalues $(3\omega)$ with tree iterations and $(-\omega)$. The eigenvectors of this transfer matrix on the column elements are obtained as $(-1,0,0,1)$, $(-1,0,1,0)$, $(-1,1,0,0)$ and $(1,1,1,1)$. According to the Heisenberg equation of motion, we have $\partial_{t}x_i(t)=[H,x_i(t)]i/\hbar$ for position and $\partial_{t}p_i(t)=[H,p_i(t)]i/\hbar$ for momentum, where $\hbar$ is the reduced Planck's constant. Therefore, the equation of motion for position and momentum can be expressed as $x_i(t)=f_1.x_i(0)+f_2.p_i(0)+\beta_t$ and $p_i(t)=f_3.x_i(0)+f_4.p_i(0)+\beta'_t$. By applying the time transformation function we get $x_i(t)=U^{\dag}_{t}x_i(0)U_t=f_1.x_i(0)+f_2.p_i(0)+\beta_t$ and $p_i(t)=U^{\dag}_{t}p_i(0)U_t=f_3.x_i(0)+f_4.p_i(0)+\beta'_t$. According to Hamiltonian Eq.(\ref{2}), the coefficients $p_i^{2}$ and $x_i^2$ form a matrix represented as $B'$. The matrix $f(t)$ is defined as follows:
\begin{equation}\label{2-2}
B'=
\left(
\begin{array}{cc}
	 0 & 1/m_i  \\
    \overrightarrow{\lambda}_i.\overrightarrow{\lambda}_ja & 0
\end{array}\right), f(t)=
\left(
\begin{array}{cc}
	f_1 & f_2  \\
    f_3 & f_4
\end{array}
\right).
\end{equation}
Where $f(t)$ applies to the ansatz function for Greens function relation $G_{t,t'}=\theta_{t,t'}f(t).f^{-1}(t')$; also $f(t)=\exp(B't)=1+B't+(B't)^2/2+(B't)^3/6+...$ is an exponential function and considering the expansion of $\sin(x)$ and $\cos(x)$, the values of $f_1$, $f_2$, $f_3$ and $f_4$ will be as follows:
\be\label{2-3}
f_1=\cos\Big(t\sqrt{\frac{\overrightarrow{\lambda}_i.\overrightarrow{\lambda}_ja}{m_i}}\Big), f_3=\sin\Big(t\overrightarrow{\lambda}_i.\overrightarrow{\lambda}_ja\Big),\nn
\ee
\be
f_2=\sin\Big(\frac{t}{m_i}\Big), f_4=\cos\Big(t\sqrt{\frac{\overrightarrow{\lambda}_i.\overrightarrow{\lambda}_ja}{m_i}}\Big),
\ee
By solving these equations of motion, position and momentum and entering the values of $f_1$, $f_2$, $f_3$ and $f_4$ were obtained as the following expressions:
\be\label{3}
x_i(t)=\cos\Big(t\sqrt{\overrightarrow{\lambda_i}.\overrightarrow{\lambda_j}a/m_i}\Big)x_i(0)+\sin(t/m_i)p_i(0)+\beta_t,\nn
\ee
\be
p_i(t)=\sin(t\overrightarrow{\lambda_i}.\overrightarrow{\lambda_j}a)x_i(0)+\cos\Big(t\sqrt{\overrightarrow{\lambda_i}.\overrightarrow{\lambda_j}a/m_i}\Big)p_i(0)+\beta'_t.
\ee
The time evolution of position and momentum is defined in terms of the time evolution function $U_t$, which gives us $\varphi_t=U_t\,\varphi_0=\exp(-H\,i/\hbar)\,\varphi_0$. Therefore, the quantum propagator of the evolving system is as follows:
\be\label{4}
x_{i}^{*}(t)\,U^{\dag}_{t}\,x_{i}(t)=x^{*}(t)\exp(H\,i/\hbar)x_i(t)=\kappa_{(x_i,t|x'_i,t_0)},
\ee
Where the symbols $x_{i}^{*}(t)$ and $U^{\dag}_{t}$ represent the complex conjugate of the functions $x_i(t)$ and $U_t$. The propagator for tetraquark systems is obtained as follows:
\be\label{5}
\kappa_{(x_i,t|x'_i,t_0)}=\frac{\exp\Big(\frac{i(x_{i}^2+x'^2_{i})\cos(t\sqrt{\overrightarrow{\lambda_i}.\overrightarrow{\lambda_j}a/m_i})}{2\hbar\sin^{-1}(t/m_i)}\Big)
}{\sqrt{(2\pi\,i\,\hbar)^4\,\sin(t/m_i)}}\exp\Big(\frac{-i}{2\hbar}\int_{0}^{t}\frac{\beta'^2_t}{m_i}dt'\Big)+\gamma,
\ee
where $\gamma=(-x_i\,x'_i)/\sin^{-1}(t/m_i)+x_i\rho_t+x'_i\tau_t$. Additionally, $\beta=\overrightarrow{\lambda_i}.\overrightarrow{\lambda_j}(\alpha_s/4)$ and $\beta'_t$ is the integral of $f(t)f^{-1}(t)\phi(t)$, where $\phi(t)=\beta\beta'$, where $f(t)$ is a $2\times2$ matrix with elements on the main diagonal $\cos(t\sqrt{\overrightarrow{\lambda_i}.\overrightarrow{\lambda_j}a/m_i})$, as well as elements on the secondary diagonal $\sin(t/m_i)$ and $\sin(\overrightarrow{\lambda_i}.\overrightarrow{\lambda_j}at)$ and $\beta'=(\pi/m_im_j)(1+2\overrightarrow{\sigma_i}.\overrightarrow{\sigma_j}/2)$, also $\sin^{-1}$ is a arcsin function. We are considering a system consisting of two identical quarks and two antiquarks that interact with each other through oscillatory interactions. The wave function of the system can be predicted by using a quantum propagator. The Feynman propagator, a function of the form Eq.(\ref{5}), can determine the subsequent times of the tetraquark system and the time evolution of the system over a time interval. Using an initial wave function input to the system as Eq.(5),
  \be\label{52}
  \varphi_{(x_i,t_0)}=N_{nl}x^{l}(\pi^2/4\alpha^2)\exp\Big(-\sum_{i=1}^{4}\alpha x_i^2\Big)Y_{lm}(x).
  \ee
 The normalized wave function  is based on the Gaussian function \cite{R17,R16,R28}. The equation for $N_{nl}$ is $(2^{l+2(2\alpha)^{\varepsilon}}/\sqrt{\pi}(2l+1))^{1/2}$, where $\varepsilon=l+3/2$. Additionally, $Y_{00}(x)=1$ and $Y_{10}(x)=\cos(x)$. This representation  is useful for decomposing position vectors. For states with a total quantum number of $l=0$,  there are representations of $[6]_{c}\bigotimes[6]_{c}$ and $[3]_{c}\bigotimes[3]_{c}$. The position part of the wave function expands on a Gaussian basis, where $\alpha$ is the variable parameter, inverely proportional to the square of the length, and $Y_{lm}$ represents spherical coordinates for spherical harmonics \cite{R29,R30}. Using the Feynman propagator $\kappa_{(x_i,t|x'_i,t_0)}$, we obtain the wave function at subsequent times $t$. The time-evolution of the initial state $\varphi(x_i;0)$ is obtained by:
\be\label{6}
\varphi_{(x_i,t)}=\int dx'_i\,\kappa_{(x_i,t|x'_i,0)}\varphi_{(x'_i,0)},
\ee
Therefore, by setting the initial wave function $\varphi_{(x_i,t_0)}$ and also inputting the quantum propagator of the system as Eq.(\ref{5}) in to Eq.(\ref{6}), and carrying out calculations and integration, the final wave function of the tetraquark system was ultimately obtained as follows:
\be\label{7}
\varphi_{(x_i;t)}=\frac{\pi^{5/2}\exp\Big(\frac{-i}{2\hbar}\int_{0}^{t}\frac{\rho_t^2}{m_i}dt'\Big) \exp\Big(\frac{ix_i^2\cos(t\sqrt{\overrightarrow{\lambda_i}.\overrightarrow{\lambda_j}a/m_i})}{2\hbar\sin^{-1}(t/m_i)}+ix_i\rho_t/\hbar\Big)}{4\alpha^2\sqrt{(2\pi\,i\,\hbar)^4\, \sin(t/m_i)}\eta'}\gamma',
\ee
where $\gamma'$ is $\exp[((-ix_i/\hbar\sin^{-1}(t/m_i))+\tau_t)^2/(-2i\cos(t\sqrt{\overrightarrow{\lambda_i}.\overrightarrow{\lambda_j}a/m_i})/\hbar\sin^{-1}(t/m_i))+4\alpha]$ and $\eta'=(\alpha-i\cos(t\sqrt{\overrightarrow{\lambda_i}.\overrightarrow{\lambda_j}a/m_i})/2\hbar\sin^{-1}(t/m))^{1
/2}$.
The wave function obtained show a complex model derived from the Hamiltonian structure. The term associated with the color interaction $\overrightarrow{\lambda_i}.\overrightarrow{\lambda_j}$ acts as a structural characteristic. This term, $\cos(t\sqrt{\overrightarrow{\lambda_i}.\overrightarrow{\lambda_j}a/m})$ is matrix-like in the wave function, and the temporal behavior of the wave function  depends on these color states, a combination of Gell-Mann matrices. When $\cos(t\sqrt{\overrightarrow{\lambda_i}.\overrightarrow{\lambda_j}a/m})$ decreases, the dynamics slow down, and the wave function changes more gradually over time, indicating  greater stability, where consistent with the theoretical understanding of QCD. The method is mass-dependent,  with increasing $m_i$ resulting in faster wave function decay, yielding a narrower, sharper graph. Furthermore, $\beta'=(\pi/m_im_j)(1+2\overrightarrow{\sigma_i}.\overrightarrow{\sigma_j}/2)$ represents the spin component of the wave function, incorporating Pauli matrices $\overrightarrow{\sigma_i}.\overrightarrow{\sigma_j}$ for spin$(s=0,1,2)$. It is worth noting that $a$ is the strength of the confinement potential with zero binding ; also the  quark-gluon coupling constant can be approximately $\alpha_s\approx1.5$ and the value of $\overrightarrow{\sigma}_i.\overrightarrow{\sigma}_j$ is obtained(for spin $s=0$ and $s=1$ in heavy quarks)\cite{R19,sigma31,si32}. The total wave function $\varphi_{(x_i,s,f,c,m_i;t)}$ includes $x,t$ for the spatial part, $s$ for the spin part, $f$ for the flavor part with $m_i$ and $c$ for the color part; all these parameters contribute to  the overall wave function.
\section{Probability Density And Dynamics Of Tetraquark Systems }\label{P}
The probability density indicates the distribution of particles in different positions and how they are present in space. By including time as a variable in the function, we can predict the dynamics and motion of the system in the future. We have calculated the probability density of the tetraquark system using the wave function of the system from Eq.(\ref{5}). The probability density is obtained as $P=\varphi_{(x_i;t)}.\varphi^{*}_{(x_i;t)}$. Therefore, we find
\be\label{8}
P_{(x_{0i};t_0)\rightarrow(x_i;t)}=\frac{(\sin^{-1}(t/m_i))^2\exp\Big(\frac{-2\alpha x_{i}^2+ 2\alpha\hbar^2(\sin^{-1}(t/m_i))^2\tau_t^2+ 2x_i\tau_t\cos(t\sqrt{\overrightarrow{\lambda_i}.\overrightarrow{\lambda_j}a/m_i})}{\cos(t\sqrt{\overrightarrow{\lambda_i}.\overrightarrow{\lambda_j}a/m_i})^2 +4\hbar^2(\sin^{-1}(t/m_i))^2\alpha^2}\Big)}{16\pi\alpha^2\hbar^2\sin(t/m_i)\Big(\cos(t\sqrt{\overrightarrow{\lambda_i}.\overrightarrow{\lambda_j}a/m_i})\Big)^2 +4\hbar^2(\sin^{-1}(t/m_i))^2\alpha^2}.
\ee
By utilizing this probability relationship and creating a probability density diagram, we can analyze how the system's behavior changes over time. This diagram illustrates the connection between observables $x_i$ and $t$, showcasing  the likelihood of locating the system within a specific interval.
The probability density $P_{(x_{0i};t_0)\rightarrow(x_i;t)}$ represents where the probability of the particles being present is concentrated at any given moment. According to the probability density relation (Eq.11), $\exp(-2\alpha x_{i}^2+ 2\alpha\hbar^2(\sin^{-1}(t/m_i))^2\tau_t^2+ 2x_i\tau_t\cos(t\sqrt{\overrightarrow{\lambda_i}.\overrightarrow{\lambda_j}a/m_i})/\cos(t\sqrt{\overrightarrow{\lambda_i}.\overrightarrow{\lambda_j}a/m_i})^2 +4\hbar^2(\sin^{-1}(t/m_i))^2\alpha^2)$, the coefficient $x_{i}^2$, as the most important part of the variance, determines the position and shows how the wave packet changes spatially. The coefficients $x_{i}$ also indicate how the center of this wave packet moves. Therefore, $P_{(x_{0i};t_0)\rightarrow(x_i;t)}\sim \exp(-2x^2/\iota(t))$ where $\iota(t)=\cos(t\sqrt{\overrightarrow{\lambda_i}.\overrightarrow{\lambda_j}a/m_i})^2 +4\hbar^2(\sin^{-1}(t/m_i))^2\alpha^2$.\\
  Hence $\Delta x(t)=\sqrt{\langle x^2\rangle-\langle x\rangle^2}=(\cos(t\sqrt{\overrightarrow{\lambda_i}.\overrightarrow{\lambda_j}a/m_i})^2 +4\hbar^2(\sin^{-1}(t/m_i))^2\alpha^2/4)^{1/2}$; for momentum $\Delta p(t)=\hbar/2\Delta x(t)=\hbar/(\cos(t\sqrt{\overrightarrow{\lambda_i}.\overrightarrow{\lambda_j}a/m_i})^2 +4\hbar^2(\sin^{-1}(t/m_i))^2\alpha^2)^{1/2}$. Thus, the Heisenberg uncertainty relation is $\Delta x(t)\Delta p(t)=\hbar/2$; meaning that no matter how much $\cos(t\sqrt{\overrightarrow{\lambda_i}.\overrightarrow{\lambda_j}a/m_i})^2 +4\hbar^2(\sin^{-1}(t/m_i))^2\alpha^2$ changes over time, uncertainty still exists and evolution is consistent with the principles of quantum mechanics.

\section{Results }

The resulting wave function provides a realistic representation of QCD, with the internal interactions investigated and the dependence of the parameters (color, flavor, and spin) on the wave function examined in the previous section. When studying the spatial oscillations of a tetraquark system over time, the parameters and degrees of freedom can be considered constant, except for the position parameter $xi$ and the time parameter $t$, and the oscillatory evolution of the system over time can be studied.
\noindent
\begin{figure}\label{F2}
	\centering
	\includegraphics[scale=0.5]
	{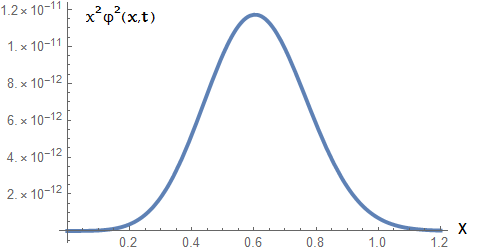}
	\caption{The probability of $\varphi_{(x_i;t)}^{2}x_{i}^2$(fm) relative to $x$(fm) is plotted at intervals $x=1.2$(fm),where time is $t=7.6$(fm/c), and is given as a variable to the wave function,(values  $t$ are on a scale of $t\times 10^{-12}$ (fm/c)).}
\end{figure}
\begin{figure}\label{F3}
	\centering
	\includegraphics[scale=0.5]
	{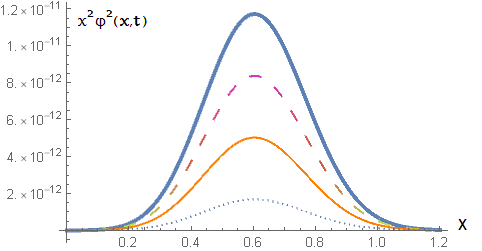}
	\caption{In this figure, it can be seen that the probability of the particle's presence increases at very short times from $t=1$(fm/c), until it reaches time $t=6$(fm/c). There are four graphs for $t=1$(fm/c)(dotted curve with the shortest peak), $t=3$(fm/c)(thin curve with the second peak), $t=5$(fm/c)(dashed curve) and $t=6$(fm/c)(thick curve), (values $t$ are on a scale of $t\times 10^{-12}$ (fm/c)).}
\end{figure}
\begin{figure}\label{F4}
	\centering
	\includegraphics[scale=0.5]
	{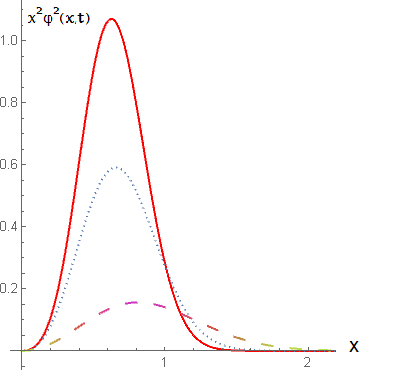}
	\caption{The function $\varphi_{(x_i;t)}^{2}x^2$(fm) is plotted as a function of $x_i$(fm) at different times $t$(fm/c). The longest wave function graph corresponds to the initial time $t_1$(fm/c), while the shortest wave function graph corresponds to $t_3$(fm/c) (for several different times), which is the final time in the evolution.}
\end{figure}
In Fig.2, we show the function $\varphi_{(x_i;t)}^{2}x^2$ based on the wave function of the tetraquark system as a function of $x$ in the range $0\lesssim x \lesssim1.2$fm for the evolution time  $t=7.6$(fm/c). The maximum value of the function $\varphi_{(x_i;t)}^{2}x^2$ is set at $1.2{\times}10^{-3}$fm when $x=0.6$fm. In Fig.3, it is evident that initially, the probability of the system increases. This indicates that the presence of the system with  two quarks and two antiquarks is reinforced in the early stages, with no indication of decline or collapse. However, as time progresses, the system moves towards decline, as illustrated in Fig.4.
The figure peak shifts towards  lower values over time, signifying system collapse. Additionally, the graph widens with time (solid-$t= 0.7(fm/c)$, dot-$t=0.5(fm/c)$ and dashed-$t=0.3(fm/c)$).\\
  In Fig.5, the system dynamics  are clearly visible. Fluctuations cause the graph to move and shift. The peak of the graph decreases from $\varphi_{(x_i;t)}^{2}x^2=1.2$fm to $\varphi_{(x_i;t)}^{2}x^2=0.07$fm, indicating a decrease. A quantum oscillation and transition between two spatial regions is observed, in which the quantum state between the two spatial states is superposed and occurs due to the kinetic energy of the expansion system. This shows that the particles (quarks and antiquarks) are not confined to a spatial boundary. Fluctuations are observable on the right side of the chart, returning to the left side over time with a lower peak. The movement  essentially oscillates back and forth  over time $t_1$. This implies that as the distance between the quark and antiquark decreases, the probability of a tetraquark's presence also decreases, with the highest probability occurring in the range  $0.5\lesssim x \lesssim0.7$(fm).\\
\begin{figure}\label{F5}
	\centering
	\includegraphics[scale=0.4]
	{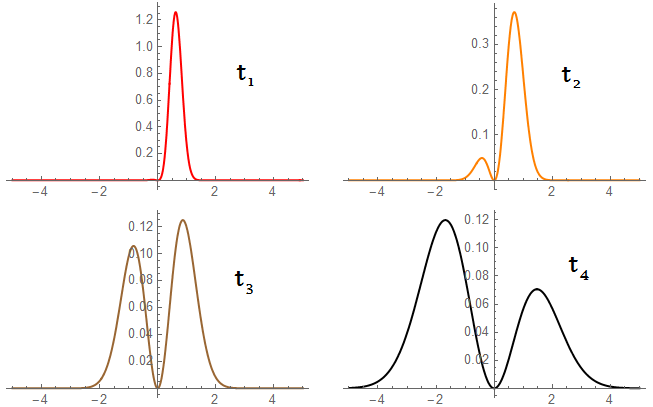}
	\caption{ The peak of each diagram moves along the x-axis as time progresses from $t_1$ to $t_4$ (in four different time periods where values are $t$ on a scale of $t\times 10^{-12}$ fm/c) during the collapse and expansion. It has shifted from $x=0.6$(fm) to $x=1.2$(fm). Additionally, the peak of the graph has collapsed and reached $\varphi_{(x_i;t)}^{2}x^2=0.07$(fm) from $\varphi_{(x_i;t)}^{2}x^2=1.2$(fm), indicating a decrease, with oscillations in the direction of propagation.}
\end{figure}
In Fig.6, the center part is brighter, indicating a higher probability of the particle's presence. As the particle approaches the darker edges, the probability of presence decreases.This density distribution provides information about the internal structure and the correlations between quarks.
The system is transitioning from an extended (delocalized) quantum state to a more localized state. That is, the probability density is concentrated at the center and the system evolves towards a stable and localized configuration in its internal degrees of freedom. Over time, as shown in Fig.5, the probability decreases and the wave function becomes broader.\\
\begin{figure}\label{F6}
	\centering
	\includegraphics[scale=0.5]
	{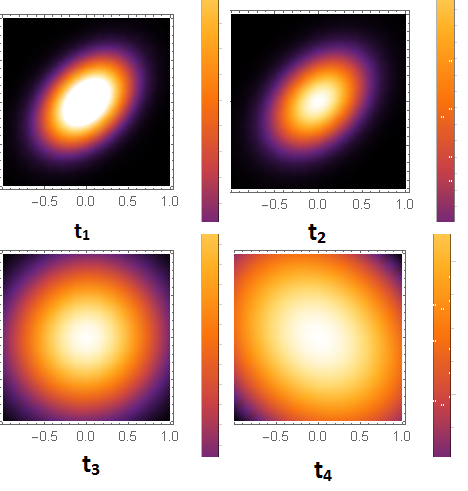}
	\caption{The probability distribution ($P_{(x_{0i};t_0)\rightarrow(x_i;t)}$) changes over time.The center part is brighter, indicating a higher probability of the particle's presence}
\end{figure}
\section{Conclusion}

In conclusion, studying the behavior of tetraquark systems is crucial for understanding their oscillations and dynamics. To comprehend the behavior of these systems and study their dynamics, it is essential to understand their wave function and how they  propagate over time, where the time evolution and oscillation patterns are examined. This was achieved, and graphs depicting the changes in behavior and collapse of tetraquark systems are shown in figures. It was observed that as the system decays, it exhibits oscillations along its axis of motion. Gluons are responsible for binding quarks together and, like an oscillator, can transfer energy and momentum between quarks and antiquarks.\\
Therefore, having a propagator and knowing the wave function of a tetraquark system with the time parameter and probability density is of great importance, which can help us solve many problems related to multiquark systems. A set of explanatory diagrams showed characteristic features of the temporal behavior of the wave function, including probability increases at early times, spatial fluctuations, diffusion, and decay-like patterns. These effects are attributed to gluon-mediated interactions, which are modeled as harmonic behavior.\\
In Figs.2-6, we have illustrated the evolution of the tetraquark system in question over time, the probability of tetraquarks existence strengthened initially but declined over time. Additionally, the oscillatory behavior of the system was clearly visible, with displacement on the propagation axis occurring simultaneously with system oscillation, as evident in the figures. This model represents the most general case applicable to all tetraquark systems necessitating the summation over all color and spin states.\\
Also, the presence of time $(t)$ as a parameter in the wave function obtained from the propagator allows for a dynamical picture of the system and shows how gluon-like couplings can affect the effective interactions in  configurations; In contrast to purely static models, including time dependence in the wave function helps to study the dynamical evolution of spatial configurations and bridges the gap between static potential models and full QCD simulations.\\
As depicted in the diagrams, the two quasi-stable states  oscillate between the two peaks of the diagram. We are  dealing with quantum oscillations between the two quasi-stable states. It is possible to exist in both regions simultaneously. These oscillations stem from the fact that gluons were considered as the oscillatory factor, behaving like an oscillator; this approach provides a step towards understanding the dynamics of entangled quarks in tetraquarks.

\bibliography{myref}

\end{document}